# 'The Boring and The Tedious': Invisible Labour In India's Gig-Economy

Pratyay Suvarnapathaki   Viral Shah   Saarthak Negi   Nimmi Rangaswamy
International Institute of Information Technology, IIIT Hyderabad

## Introduction and Background

India's burgeoning gig economy, particularly the food delivery sector spearheaded by platforms like Swiggy and Zomato, presents a complex socio-technical landscape. It is an environment characterised by rapid digital adoption driven by economic necessity, yet marked by deep-rooted structural inequities and infrastructural challenges (Baa, 2023; Midha, 2016). A generations-spanning wave of Indians has joined the digital sphere, comprising the 'Next Billion Internet Users', demonstrating functional digital literacy acquired primarily through the imperative of earning a livelihood (Google LLC, 2021; Kumar, 2017; Medhi Thies, 2015). However, this adaptation often masks significant systemic friction between the streamlined interfaces of delivery applications and the messy, unpredictable realities of on-the-ground labour conditions (Bejaković, 2020; Nikore, 2021; Straughan & Bissell, 2022).

While much research, particularly from the Global North, concentrates on algorithmic exploitation and precarious work conditions inherent in gig labour (Graham, 2018; Lang et al., 2023; Lee, 2018), the Indian context reveals a more nuanced narrative as these new digital platforms undeniably offer crucial income streams (Fairwork India, 2024), often serving as a vital economic lifeline for individuals from marginalised communities. Our study diverges from perspectives that either solely critique exploitation or valorise opportunity, instead choosing to centre the lived experiences and voices of the delivery workers themselves. Through grounded, field research we uncover cycles of invisible labor (Gilbert, 2023) that contribute to a overarching phenomenon of 'digital discomfort' – the chronic, everyday friction experienced by workers as they navigate the gap between the app's simplified, often gamified, representation of work and the inherent chaos and complexities of real-world logistics and urban environments.

## Algorithmic Management and Two-Fold Gamification

Delivery platforms extensively utilise algorithmic management techniques and gamification strategies to direct, monitor, and incentivise worker productivity. Features like Zomato's medal tiers or Swiggy's incentive zones are designed to nudge workers towards platform-defined goals, often framed as opportunities for higher earnings. However, workers are far from passive subjects within this system. They actively devise and deploy creative counter-strategies and workarounds to reclaim a degree of control, optimise earnings within imposed constraints, and mitigate systemic inefficiencies.

This interactive dynamic, where both platform and worker engage in strategic manoeuvres, has been described as 'two-fold gamification' (Goyal, 2024). It underscores the agency workers exercise within tightly constrained digital environments, while simultaneously casting the ethical tensions of platform design into sharp relief. For instance, a common tactic observed is delivery partners marking themselves as 'reached' at a restaurant or customer location prematurely. This action is often a calculated move to satisfy the app's metrics and avoid potential (often opaque) penalties for perceived delays, even if the delay originates from the restaurant or unsuccessful attempts to reach difficult-to-locate addresses on time.. Such strategies are born from significant representational gaps – the discrepancies between the app's portrayal of the delivery process (e.g., order readiness times, estimated travel times) and the actual ground reality (e.g., kitchen delays, traffic congestion, inaccurate map data). Furthermore, the constant barrage of notifications, status update requirements, and mandatory interactions contributes to significant notification fatigue and cognitive load (Lang et al., 2023), pushing workers to find shortcuts merely to cope with the digital demands of the job.

## Methodology

To investigate the nature and extent of invisible labour and digital discomfort, we adopted a qualitative, worker-centred research approach. Our primary data collection method involved semi-structured interviews conducted with 14 food delivery workers operating in Hyderabad, India. These workers were engaged with major platforms prevalent in the area, including Swiggy, Zomato, Blinkit, and BigBasket. Participants were recruited using convenience sampling, a pragmatic choice dictated by the challenges of accessing this mobile and often transient workforce. Despite this limitation, the sample encompassed a range of experience levels, from workers who had been on the platforms for approximately 6 months to veterans with up to 9 years in the delivery sector.

The interviews were designed to elicit rich, detailed accounts of the workers' daily routines, focusing specifically on their interactions with the delivery applications, digital and physical challenges, and strategies to navigate platform rules and real-world obstacles. We probed into specific aspects of their workflow, including order acceptance, navigation, communication with restaurants and customers, dealing with delays, understanding earnings and incentives, and managing the app interface itself. Thematic analysis was employed to identify recurring patterns and key issues across the interviews, allowing us to surface the core components of invisible labour as experienced by the workers. This approach prioritised capturing the nuances of their lived experiences, ensuring that our findings were grounded in their perspectives rather than solely relying on platform-centric views of efficiency.

| P | 1 | 2 | 3 | 4 | 5 | 6 | 7 | 8 | 9 | 10 | 11 | 12 | 13 | 14 |
|---|---|---|---|---|---|---|---|---|---|---|---|---|---|---|
| Age | 36 | 19 | 20 | 44 | 25 | 25 | 23 | 26 | 32 | 43 | 25 | 21 | 33 | 28 |
| Gig-Work Tenure | 9 years | 6 months | 1 year | 3 years | 3 years | 1 year | 4 years | 9 months | 1 year | 1 year | 1 year | 2 years | 4 years | 1 year |
| Platforms Worked | Shadowfax, Swiggy | Blinkit | Zomato, Swiggy, Shadowfax | Bigbasket | Zomato | Zomato | Swiggy, Zomato | Swiggy | Zomato | Swiggy | Swiggy, Rapido | Swiggy | Zomato, Swiggy | Zomato |
| Shift Duration (hours) | 10.5 | 13 to 14 | Variable part-time | Variable part-time | 11 | 12 | 11 to 12 | 10 to 12 | 10 | 5 to 6, part-time | 10-12 | 8-10 | 8-10 | Variable part-time |

Table: Participant Information & Demographics

## Findings: The Invisible Labour of Waiting and Repetition

Our analysis, grounded in the detailed narratives of the 14 delivery workers interviewed, identified two predominant forms of uncompensated, yet essential, labour inherent in their daily work. Often rendered invisible by the platform's focus on completed deliveries, these tasks constitute a significant burden and contribute heavily to digital discomfort. We term this phenomenon 'invisible labour':

1. Pervasive Waiting Time: A substantial portion of a delivery worker's shift is consumed by waiting – estimates from participants suggest this can account for approximately 25-30% of their active time on the platform. This waiting occurs in multiple contexts:
    - At Restaurants: Workers frequently arrive at restaurants only to find the order delayed or pending, despite app notifications suggesting otherwise. This discrepancy stems from poor integration between platform systems and actual kitchen workflows, or restaurants potentially deprioritising delivery orders. Workers described being forced to wait idly, sometimes in designated areas outside the main restaurant space, feeling disrespected and unproductive.
    - Between Orders: Significant downtime occurs while waiting for the platform's algorithm to assign the next order. This is often exacerbated in oversaturated markets where the number of active riders exceeds immediate demand, or in areas misleadingly marked as 'high-demand zones' which fail to yield consistent orders. The lack of predictability in order allocation adds a layer of anxiety to this waiting period.
    - Post-Delivery: Waiting for customers who may be slow to respond at the drop-off location, navigating complex apartment buildings, or dealing with security personnel at gated communities also contributes to uncompensated time. Issues like inaccurate or outdated addresses further prolong this process, often requiring calls to the customer for clarification.

2. The root causes identified for this excessive waiting include market oversaturation, opaque and unpredictable algorithmic order assignment, miscommunication facilitated by the app (like incorrect 'order ready' statuses or unmarked addresses ), and logistical challenges inherent in urban environments. Crucially, this waiting time is generally not factored into the workers' compensation structure.

3. Tedious Repetitive Tasks: Beyond waiting, workers endure a significant amount of repetitive, low-value interaction work mandated by the platform interfaces. These tasks, while individually small, collectively contribute to cognitive strain and operational inefficiency:

    - Constant App-Switching: Workers frequently toggle between the delivery platform's app and external navigation apps like Google Maps because the in-built navigation often lacks real-time traffic updates or efficient routing. This constant switching is mentally taxing and drains the phone battery.
    - Redundant Updates & Interactions: The workflow demands multiple, often seemingly redundant, status updates (e.g., 'reached restaurant', 'order picked up', 'reached customer'). Additionally, mandatory actions like taking photos of delivered orders add to the repetitive burden.

- Repetitive Communication: Workers often find themselves making routine calls to customers just before arrival, a task some wish could be automated. Dealing with customer support for issues like incorrect addresses or order cancellations is also frequently cited as a frustrating and time-consuming repetitive process, often perceived as unsupportive towards the worker.

4. These interaction overheads stem largely from platform designs that prioritise data collection and metric tracking over the delivery agent's experience and cognitive comfort. The cumulative effect is not just lost time, but increased mental fatigue and frustration, directly impacting the worker's experience and potentially their safety (e.g., handling communications while driving).

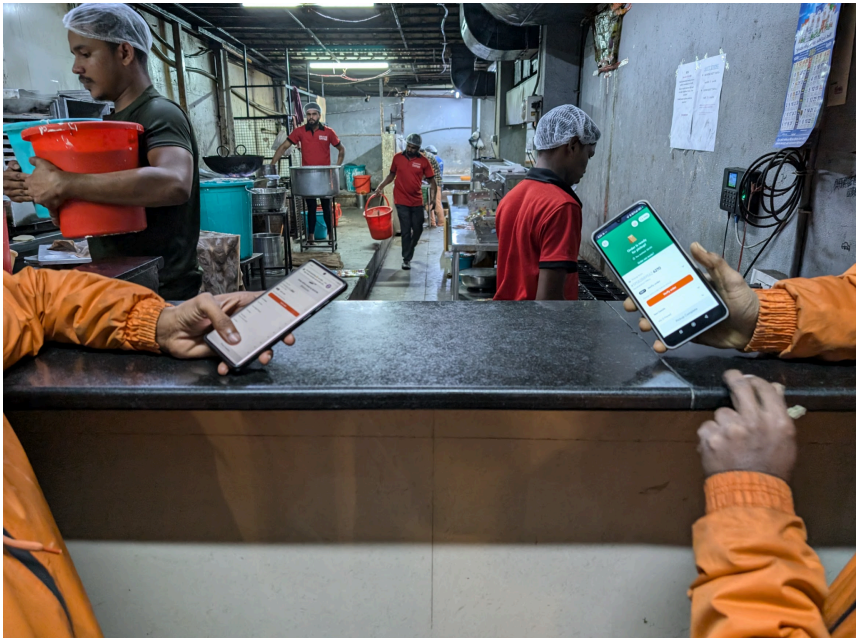

Figure: Two workers (interview participants) waiting to collect orders at a popular restaurant's takeout counter

## Future Scope: Worker-Centred Automation

To address the invisible labour of waiting and repetitive digital tasks, worker-centred automation presents a promising way to enhance digital comfort without displacing workers. GUI automation tools such as AutoDroid (Wen et al., 2024) and conceptual systems like GigSense (Imteyaz, 2024) could help streamline routine interactions, including status updates, verification steps, app-switching for navigation, and basic communication with customers.

While many workers are open to automating tedious aspects of their work, our findings reveal a degree of scepticism. Concerns include increased platform control, surveillance, and the limited ability of automation to address challenges in the physical world. Therefore, while co-designing platforms with worker inputs (Imteyaz, 2024) may be more ideal than consistently attainable, sustained engagement through immersive research into gig work remains essential. This approach ensures that AI automation aligns with labour-centric practices, preserves worker agency, adapts to local contexts, and avoids reinforcing platform-centric control. The goal should remain on meaningful augmentation rather than replacement.

Two promising design directions include:

1. Ambient automation, which minimises friction by working in the background without requiring constant interaction.

2. Customisable micro-automations which give workers the ability to choose and tailor task-level shortcuts based on their individual preferences and routines.

## Conclusion: HCI for Equitable Futures

The experiences of Indian food delivery workers highlight a need for HCI in the Global South to prioritise 'digital comfort' over narrow efficiency metrics that cause quantification bias. Designs must align with workers' lived realities, mitigating invisible labour. This involves reframing automation not for surveillance but as a tool for digital workerism (Calacci & Pentland, 2022), augmenting capabilities and agency. Technical interventions (UI design, algorithms, automation) must bridge representational divides, increase transparency, improve usability, and amplify worker agency to build humanely sustainable platforms.